\documentclass[submission,copyright,creativecommons,hidelinks]{eptcs}
\providecommand{\event}{TFPIE 2021} 
\usepackage{underscore}           
\usepackage{url}
\usepackage{graphicx}
\usepackage{stackengine}
\usepackage[square,sort,comma,numbers]{natbib}
\usepackage[toc,page]{appendix}

\usepackage{xcolor}
\definecolor{hyperlinks}{HTML}{1000E4}

\title{Teaching Programming to Novices \\ Using the codeBoot Online Environment}
\author{Marc Feeley and Olivier Melançon
\institute{Université de Montréal\\ Montréal, Canada}
\email{feeley@iro.umontreal.ca \quad olivier.melancon.1@umontreal.ca}
}

\def\authorrunning{M. Feeley, O. Melançon}
\def\titlerunning{Teaching Novices with codeBoot}

\begin{document}

\maketitle

\begin{abstract}
  Teaching programming to novices is best done with tools with simpler
  user interfaces than professional IDEs that are tailored for
  experienced programmers.  In a distance learning situation it is
  also important to have a development environment that is easy to
  explain and use, and that integrates well with the variety of course
  material used (slides, homework, etc).  In this paper we give an
  experience report on teaching programming with codeBoot, an online
  programming environment we designed specifically for novices.
\end{abstract}

\section{Introduction}

The transition from in-person to online teaching of computer
programming is challenging both for students and professors who must
find new ways to interact effectively.  In this paper we share our
experience teaching a University level first programming course during
the Fall 2020 semester which, like in most places around the world,
was done through distance learning.  To facilitate this we designed
codeBoot, an online programming environment geared towards novices
with no prior programming experience.  This environment supports the
JavaScript and Python languages and thus allows teaching imperative
programming concepts as well as functional programming concepts.  We
discuss both our experience teaching with codeBoot as the main
technical tool and the design and implementation of codeBoot.

\subsection{Course Outline}

\emph{Programmation 1} is a mandatory programming course in our computer
science undergraduate curriculum.  The topics covered are fairly
typical for a first programming course and includes basic types, arrays
and records, loops, structured programming, procedural abstraction,
test driven development, web programming and event driven execution.
We notably do not teach object oriented programming, which is covered
in the programming course that comes next in the curriculum, but we do
show various basic aspects related to functional programming,
including recursion, higher-order functions (\verb|map| and
\verb|reduce|) and callbacks.

The course was previously taught for several years using the
JavaScript language and the Fall 2020 semester was the first time
using Python.  The transition to Python was mostly motivated by a
desire to give the students experience in a language that will be used
for other courses in the following semesters.\footnote{Our department
  has a large machine learning group and Python is the language of
  choice there.}  We strive to teach concepts of programming that
apply to most languages rather than teaching Python idioms.  The
intent is to enable the students to quickly adapt their knowledge to
other programming languages once the course is over.  In particular we
use very few of the standard modules and Python specific features,
preferring to show how (sometimes complex) things can be programmed
from basic constructs, which is not only a pedagogical demonstration
of the power of abstraction but a skill that all competent programmers
should master.

The course spans 13 weeks and there are 6 hours of virtual contact per
week with the students: 3 hours of lectures with the professor and 3
hours of virtual \emph{lab time} with a teaching assistant.  There are
10 short programming homeworks and 2 larger scale project (500 to 1000
lines of code each). The virtual contact with the students is done
with standard videoconferencing software and there is an online portal
for accessing a question/answer forum and various documents including
slides and video recordings for all the lectures.  To minimize
software version related problems, most critically during online
quizzes and exams, we ask that students install the latest version of
the Firefox, Chrome or Safari browsers.

\subsection{Student Profile}

Enrolment was 265 students of diverse backgrounds.  A minority had
previous experience in programming, either because they are
self-taught or because they have taken a programming course that did
not meet the criteria for course equivalency, such as a high-school
level course.  Some students also came with a limited mathematical
background.  Finally, their computing habits are typically narrowly
focused on a single operating system and ecosystem.  The majority have
Microsoft Windows installed on their computer and few know about
GNU/Linux, the concepts of a \emph{shell} and \emph{command line
  interface}, and don't have the skills to replace their operating
system or install and configure a virtual machine.

\section{Design of codeBoot}

Over the past years teaching \emph{Programmation 1} we have come to
the realization that professional-grade programming tools, either based
on command line tools or IDEs that must be installed and
configured, are far from ideal for teaching novices. 
The typical user interface (UI) is overwhelming for novices and a distraction from
learning the essence of programming, which is how to write correct,
well structured and maintainable code. In \hyperref[design:pycharm]{Figure~\ref{design:pycharm}}, we show the PyCharm \cite{jetbrainsPyCharm2020} user interface, one of the most popular Python IDEs. Apart from the \textit{run} and \textit{debug} buttons, PyCharm also offers over 40 other clickable buttons, including options to execute the code with coverage, profile the code and manage Python virtual environments. These features can be confusing to novices and are out of the scope of our course.

Moreover, such tools have installation and usage procedures that are error prone and that depend on the type and version of the operating system. In the case of Python, multiple versions tend to be installed and managed with the help of virtual environments. This causes unnecessary delays in the learning process and puts students at risk of unknowingly using different versions. This issue can be mitigated for in-person teaching by setting up labs of machines with
identical software. For remote teaching, not only is this not
possible but it is hard to explain to students how to install a
programming environment on their computer due to the wide variability
and not being able to assist the students in ``hands on'' sessions.

For this reason we have designed our own programming environment,
codeBoot, with a UI geared towards novices.  It requires no
installation by running entirely inside the student's web browser
either on a desktop or mobile device.  The earlier versions supported the
JavaScript language and for the Fall 2020 semester we extended it to
also support Python.  The interface between the UI and the language
now allows new languages to be added relatively easily (more on this
in Section~\ref{sec:implementation}). For example we are currently adding support for the Scheme language.

\subsection{Single Stepping}

\begin{figure*}[ht]
  \centering
  \stackanchor{
    \begin{minipage}{3.0in}
      \href{https://codeboot.org/py}{\includegraphics[width=3.3in]{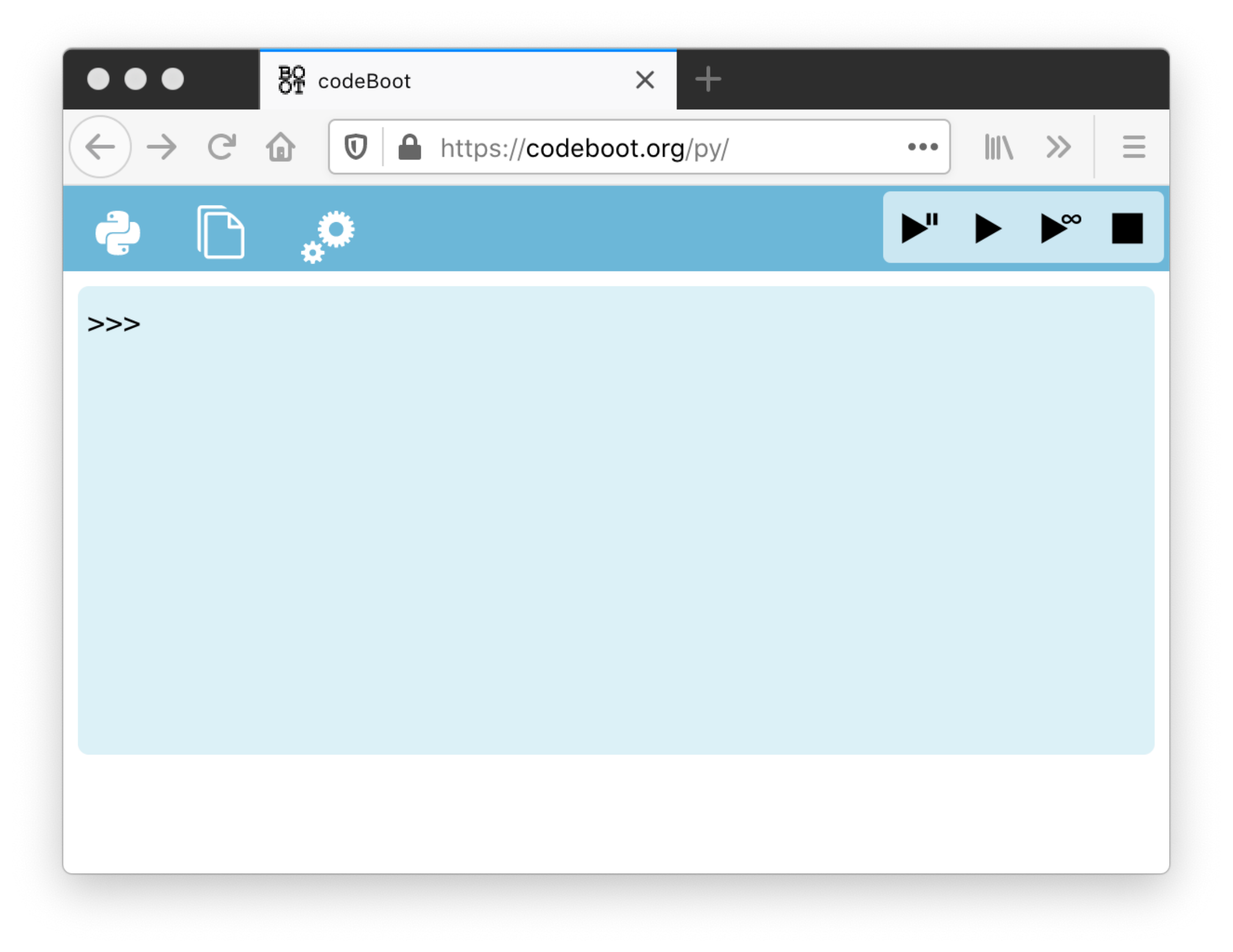}}
    \end{minipage}
    }{~~~~~~~~~~(a) Appearance on first visit of codeBoot}
  \stackanchor{
    \begin{minipage}{3.0in}
      \href{https://codeboot.org/py/?init=WCxm1UJ8iylrYuwbaHjacSYMt9dhxSi-61CChbhYzqpZVsGgDgOzaCPDLQcCNTg6vghdOS0Oh78culi9Pt8NhH0dP9S8XdcDuMXtlIXSLcxGfg6vs4rFNzYsFVvh4vyH,ibnVtYmVycyA9IFswLCAxMCwgMjAsIDMwXQ==,e,ibGlzdChtYXAoYm9vbCwgbnVtYmVycykp,e,ibGlzdChtYXAobGFtYmRhIHg6IHgqKjIsIG51bWJlcnMpKQ==,e14}{\includegraphics[width=3.3in]{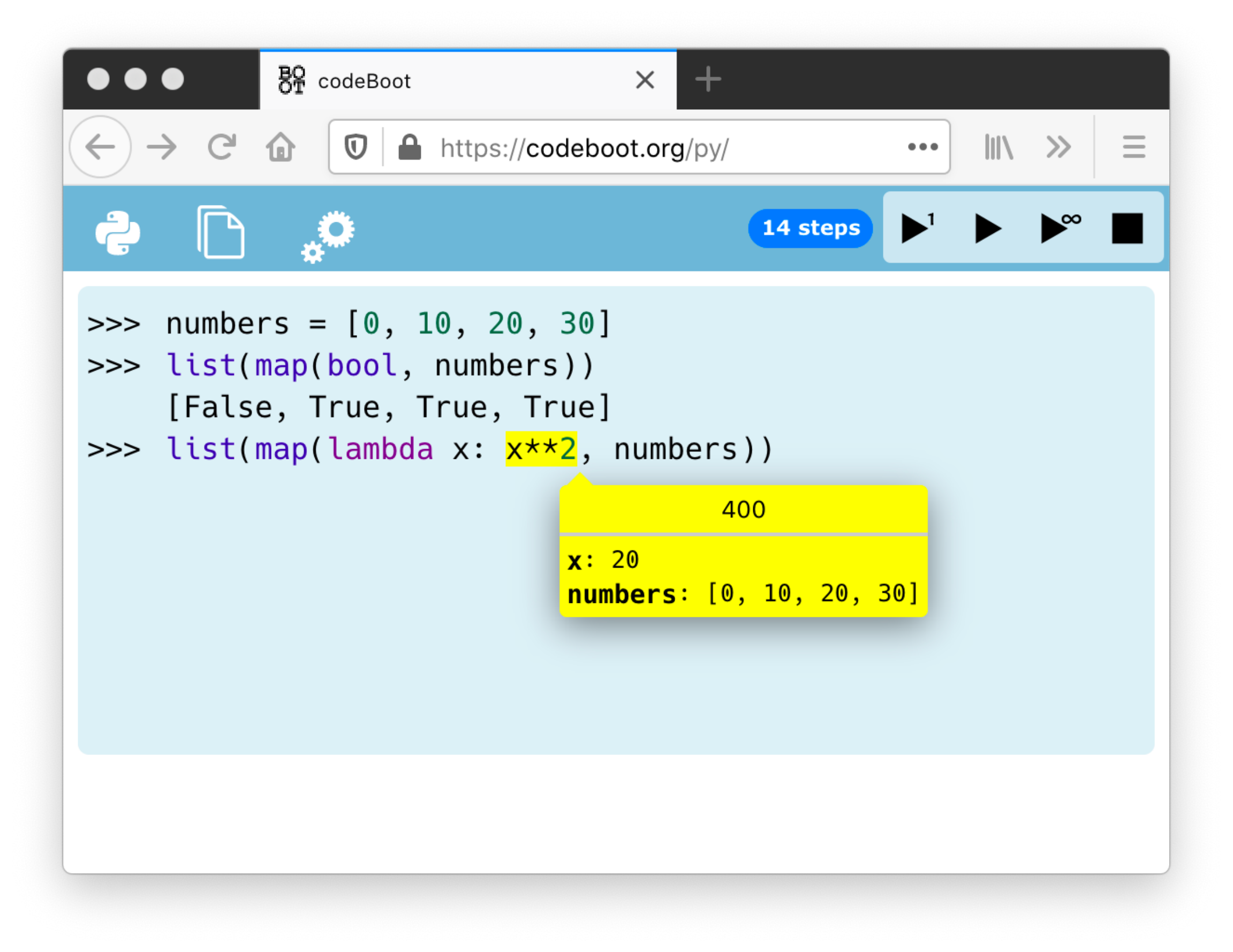}}
    \end{minipage}
    }{~~~~~~~~~~(b) Single stepping at the REPL}
  \begin{minipage}{6.0in}
    \vspace*{0.7in}
  \end{minipage}
  \stackanchor{
    \begin{minipage}{3.0in}
      \href{https://codeboot.org/py/?init=YXzcxegSqsBFJCvZ4CDWcCZi19awsjU9DVvglid-Y8ecSmpzAmjDpXXNaAmxfx-pPc7fHrFMFLQ3bDMRnTtE0kiDy9dPg_BcbZlVw-Qmi7vcIi4KnSPFFA0aZwCUFgo7,Fc3BpcmFsLnB5,ZGVmIHNwaXJhbChuKToKICAgIGlmIG4gPiAwOgogICAgICAgIGZkKG4pOyBsdCg5MCkKICAgICAgICBzcGlyYWwobi01KQoKY2xlYXIoMjUwLCAxMDApOyBnb3RvKDAsIC00MCk7IHNwaXJhbCg4MCk=,e112}{\includegraphics[width=3.3in]{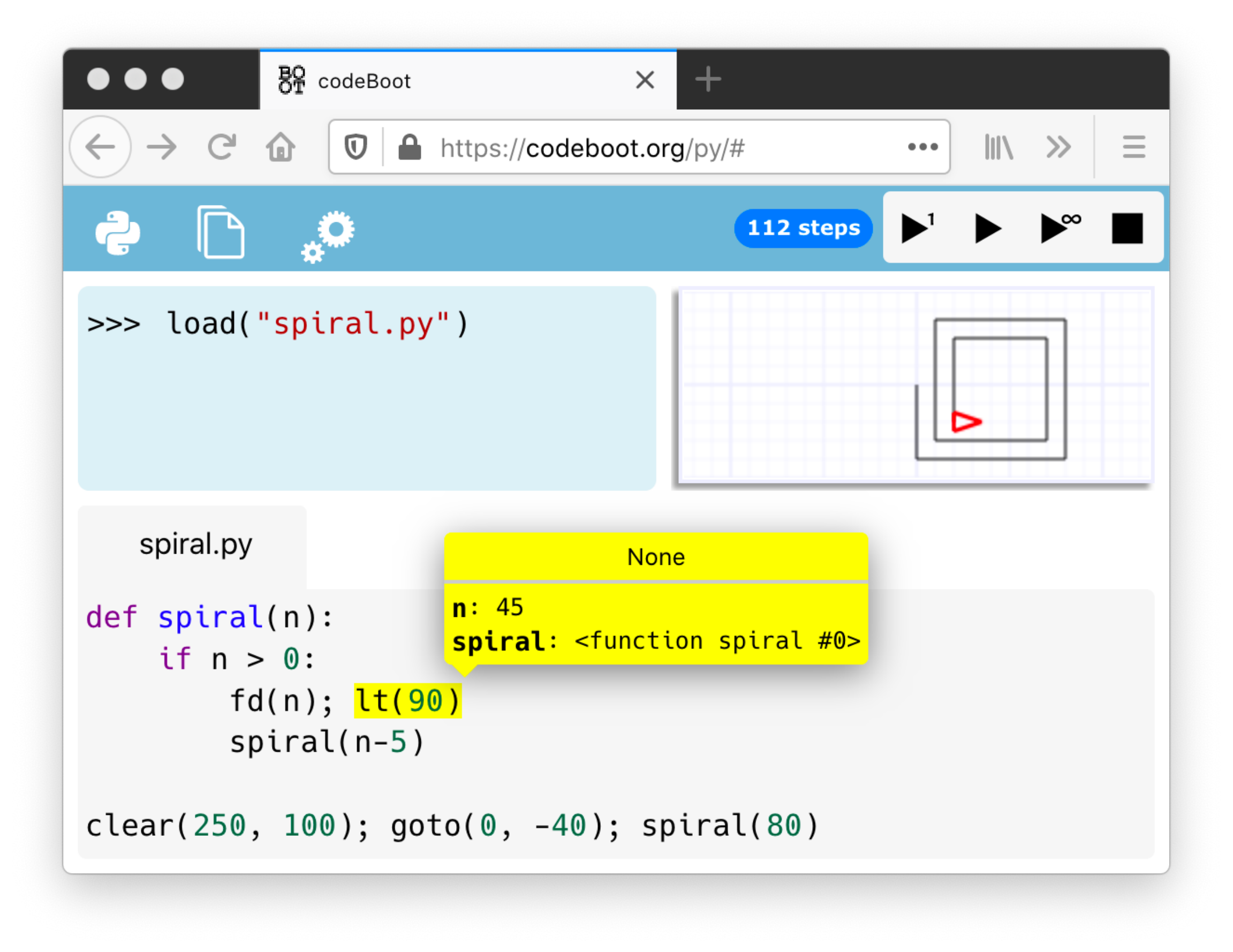}}
    \end{minipage}
    }{~~~~~~~~~~(c) Program drawing with the turtle}
  \stackanchor{
    \begin{minipage}{3.0in}
      \href{https://codeboot.org/py/?init=A-3jCOlaM3H-SJ3xhYEwsYchQezrRivu4XzrtfqiLF7HHklDtWpvC2kzooifa2mGG88TNjx7xRZd3D_YXj2_PSx9T-hYPWuOj6-Ai_JqS5vhgqZVx3t0r9oLU4wHfnuj,FZXZlbnRzLnB5,d2luID0gZG9jdW1lbnQucXVlcnlTZWxlY3RvcignLmNiLWh0bWwtd2luZG93JykKd2luLmlubmVySFRNTCA9ICc8YnV0dG9uIG9uY2xpY2s9ImNsaWNrKCkiPkFERCAxPC9idXR0b24-JwoKY291bnQgPSAwICAjIGNvdW50IG9mIHRoZSBudW1iZXIgb2YgY2xpY2tzIG9mIHRoZSBidXR0b24KCmRlZiBjbGljaygpOiBnbG9iYWwgY291bnQ7IGNvdW50ICs9IDE7IHByaW50KGNvdW50KQ==,e}{\includegraphics[width=3.3in]{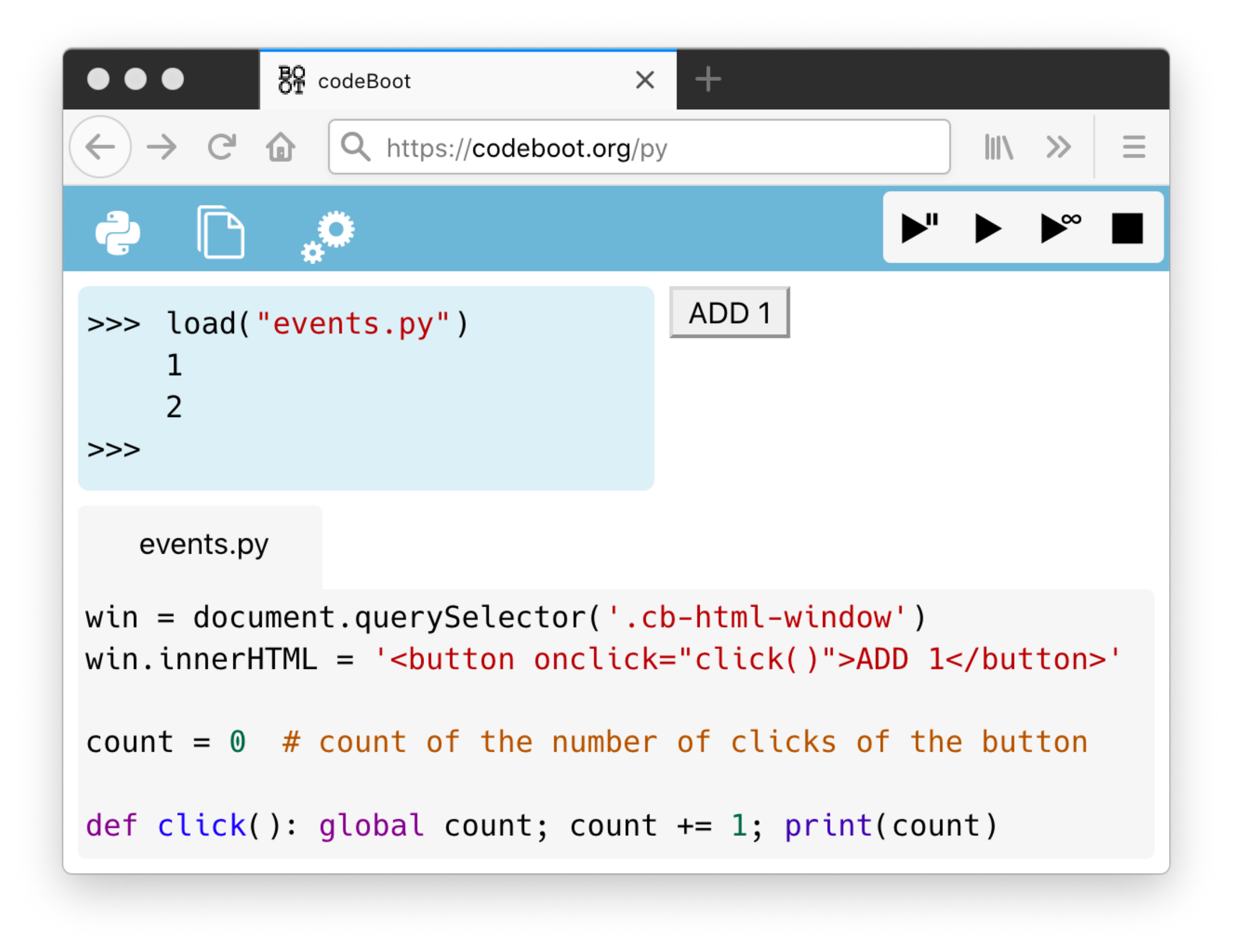}}
    \end{minipage}
    }{~~~~~~~~~~(d) Program handling DOM events}
  \caption{The codeBoot user interface and its features (each screenshot is a clickable hyperlink)}\label{fig:codeboot-ui}
\end{figure*}

The main pedagogical feature of codeBoot is support for single
stepping the execution.  We find that this is the best way for novices to
understand the details of the programming language's execution model:
how values are computed (data flow) and how operations are chained
(control flow).  When debugging more complex programs the single
stepping helps the students understand the faulty behavior of their
programs \cite{xinogalosMicroworldsGamesAnimations2017}.  It partially makes up for the absence of in-person teaching
assistants to explain unexpected behaviors to the students.

Figure~\ref{fig:codeboot-ui} shows screenshots of codeBoot's UI in
various situations.  The UI is intentionally kept to a bare minimum.
As can be seen in
\href{https://codeboot.org/py}{\color{hyperlinks}{Figure~\ref{fig:codeboot-ui}a}}
the initial appearance has only three parts: four buttons to control the
execution of code in the upper right (single step, execution by time
lapse repetitive steps, execution until the end, and stop execution),
a set of three menus in the upper left (language selection, local file
creation/selection, and preferences like speed of time lapse execution
and font size), and a console in which code can be entered and
executed using a Read-Eval-Print-Loop (REPL).  Early in the course,
the REPL allows quick experimentation of code execution.  Later on, it
allows inspecting the program state when debugging.  When local files
are created they appear in tabs at the bottom of the window.  The
files managed by codeBoot are local to the browser with no link to the
host operating system's file system.  However, files can easily be
copied in and out of the browser with a drag-and-drop operation, and
files persist within the browser over successive codeBoot sessions.

To display the state of the program when single stepping
(Figures~\href{https://codeboot.org/py/?init=WCxm1UJ8iylrYuwbaHjacSYMt9dhxSi-61CChbhYzqpZVsGgDgOzaCPDLQcCNTg6vghdOS0Oh78culi9Pt8NhH0dP9S8XdcDuMXtlIXSLcxGfg6vs4rFNzYsFVvh4vyH,ibnVtYmVycyA9IFswLCAxMCwgMjAsIDMwXQ==,e,ibGlzdChtYXAoYm9vbCwgbnVtYmVycykp,e,ibGlzdChtYXAobGFtYmRhIHg6IHgqKjIsIG51bWJlcnMpKQ==,e14}{\color{hyperlinks}{\ref{fig:codeboot-ui}b}}
and
\href{https://codeboot.org/py/?init=YXzcxegSqsBFJCvZ4CDWcCZi19awsjU9DVvglid-Y8ecSmpzAmjDpXXNaAmxfx-pPc7fHrFMFLQ3bDMRnTtE0kiDy9dPg_BcbZlVw-Qmi7vcIi4KnSPFFA0aZwCUFgo7,Fc3BpcmFsLnB5,ZGVmIHNwaXJhbChuKToKICAgIGlmIG4gPiAwOgogICAgICAgIGZkKG4pOyBsdCg5MCkKICAgICAgICBzcGlyYWwobi01KQoKY2xlYXIoMjUwLCAxMDApOyBnb3RvKDAsIC00MCk7IHNwaXJhbCg4MCk=,e112}{\color{hyperlinks}{\ref{fig:codeboot-ui}c}}),
codeBoot highlights in the code the expression whose evaluation has
just finished and uses an environment bubble attached to that location
that contains the set of variables that are in scope and their value.
For simple programs, with relatively few variables, this gives a clear
picture of the program's state.  Control statements such as
\texttt{if}, \texttt{while}, \texttt{for}, \texttt{try},
\texttt{return}, \texttt{raise}, and \texttt{assert} that have an
expression in their syntax do not count as a step because highlighting
the expression is sufficient to track the flow of execution, and
avoids \emph{micro steps} that have redundant information.  The
statements \texttt{pass}, \texttt{break}, \texttt{continue},
\texttt{return} (with no value), \texttt{raise} (with no value),
\texttt{def}, \texttt{class}, and assignment are counted as a step and
are highlighted with a bubble. The later three indicate in the bubble
the value being assigned and the destination.

The single stepping mechanism keeps track of the number of steps and
displays a counter next to the single step button.  This is
particularly useful to convey a sense of execution cost and to address
the topics of program optimization and algorithmic complexity (the
course skims these topics to give an informal understanding).

\subsection{Playground and Web Applications}
\label{sec:webapp}

To keep students engaged in the learning process it is important to
have them write programs that implement modern forms of user
interaction that go beyond textual input/output at the console \cite{vidalduarteTeachingFirstProgramming2016} (or
calls to the browser's \texttt{alert}, \texttt{prompt} and
\texttt{confirm} functions that are also supported by codeBoot).  For
this purpose codeBoot provides a \emph{playground} area that appears
on the right of the console and that enables one of the following three
types of user interactions.

\begin{itemize}
\item Drawing pictures using the turtle metaphor by calling builtin
  functions compatible with Python's standard \texttt{turtle} module
  as shown in \href{https://codeboot.org/py/?init=YXzcxegSqsBFJCvZ4CDWcCZi19awsjU9DVvglid-Y8ecSmpzAmjDpXXNaAmxfx-pPc7fHrFMFLQ3bDMRnTtE0kiDy9dPg_BcbZlVw-Qmi7vcIi4KnSPFFA0aZwCUFgo7,Fc3BpcmFsLnB5,ZGVmIHNwaXJhbChuKToKICAgIGlmIG4gPiAwOgogICAgICAgIGZkKG4pOyBsdCg5MCkKICAgICAgICBzcGlyYWwobi01KQoKY2xlYXIoMjUwLCAxMDApOyBnb3RvKDAsIC00MCk7IHNwaXJhbCg4MCk=,e112}{\color{hyperlinks}{Figure~\ref{fig:codeboot-ui}c}}.
  There is also a \texttt{getMouse()} builtin function to get the
  location of the mouse and the state of the mouse buttons and
  \texttt{shift}/\texttt{ctrl}/\texttt{alt} keys (a program
  to do freehand drawings with the mouse is a mere
\href{https://codeboot.org/py/?init=VCDgjydf6gWJP40mN_6YTAefo8_2ftB37-kiZ5EuFxRnbbuNzNCnzXJixStEwrS-1nfhtNxXsKtcBv4yYxJeQzDs4PCCjPrI3pOa3grbg2jaGKfjEP2427rm7dI_tONX,showLineNumbers:true,FZG9vZGxlLnB5,IyBGaWxlOiBkb29kbGUucHkKCmh0KCk7IHB1KCkgICMgc3RhcnQgd2l0aCBwZW4gdXAgYW5kIGhpZGRlbiB0dXJ0bGUKCndoaWxlIFRydWU6CiAgICBtID0gZ2V0TW91c2UoKQogICAgaWYgbS5idXR0b246IHBkKCkgICMgYnV0dG9uIHByZXNzZWQgc28gZHJhdyBhIGxpbmUKICAgIGdvdG8obS54LCBtLnkpCiAgICBpZiBtLmJ1dHRvbjogcHUoKSAgIyBzdG9wIGRyYXdpbmcKICAgIHNsZWVwKDAuMDUpICAgICAgICAjIGRvbid0IHVzZSB0b28gbXVjaCBDUFU=,e}{\color{hyperlinks}{10 lines of code}}).
\item Drawing pictures using a simulated screen (rectangular grid of
  pixels).  Each pixel's color can be controlled with the
  \texttt{setPixel(\textit{x},\textit{y},\textit{color})} builtin function.
  The \texttt{getMouse()} function
  also works with the simulated screen but reports pixel grid
  coordinates.  The simulated screen is appropriate for implementing
  some types of video games and to make
  \href{https://codeboot.org/py/?init=jKTy51rDi7YzjI4v--Rj5H2R_QMKdgUGaaGNd6fZPznUU_CVKLDby6YqKEFZL6Qmqipzt7Kq2AiznkRuF59lCDrGhG0ka87JlunOnOJ_7z6ZFgOTQJiyOhXyabhlI3UD,FcmFuZG9tLXBpeGVscy5weQ==,IyBUaGlzIHByb2dyYW0gZmlsbHMgYSAxOHgxMiBncmlkIG9mIHBpeGVscyB3aXRoIHJhbmRvbSBjb2xvcnMuCiMgQXQgdGhpcyBsb3cgcmVzb2x1dGlvbiBjb2RlQm9vdCByZXByZXNlbnRzIGVhY2ggc2ltdWxhdGVkIHBpeGVsCiMgd2l0aCBzZXZlcmFsIHBpeGVscyBvbiB0aGUgcmVhbCBzY3JlZW4uICBUaGUgcmVzb2x1dGlvbiBpcyBzZXQKIyB3aXRoIHRoZSBmdW5jdGlvbiBzZXRTY3JlZW5Nb2RlKHdpZHRoLGhlaWdodCkuICBUaGUgY29sb3Igb2YgYQojIHBpeGVsIGlzIHNldCB3aXRoIHNldFBpeGVsKHgseSxjb2xvciksIHdoZXJlIHRoZSBjb2xvciBpcyByZXByZXNlbnRlZAojIGJ5IGEgc3RydWN0dXJlIGNvbnRhaW5pbmcgdGhlIFJHQiBjb21wb25lbnRzIG9mIHRoZSBjb2xvci4gIEVhY2gKIyBjb21wb25lbnQgaXMgYW4gaW50ZWdlciBmcm9tIDAgdG8gMTUuICBUaGUgY29udGVudCBvZiB0aGUgc2NyZWVuIGNhbgojIGJlIGNvbnZlcnRlZCB0byBhIHN0cmluZyB3aXRoIHRoZSBleHBvcnRTY3JlZW4oKSBmdW5jdGlvbi4gIFRoaXMgY2FuCiMgYmUgdXNlZCBmb3IgdW5pdCB0ZXN0aW5nIGNvZGUgdGhhdCBtb2RpZmllcyB0aGUgc2NyZWVuLgoKbnggPSAxMgpueSA9IDgKCmRlZiByYW5kMTYoKToKICAgIHJldHVybiBpbnQocmFuZG9tKCkqMTYpCgpzZXRTY3JlZW5Nb2RlKG54LCBueSkKCmZvciB5IGluIHJhbmdlKG55KToKICAgIGZvciB4IGluIHJhbmdlKG54KToKICAgICAgICBjID0gc3RydWN0KHI9cmFuZDE2KCksIGc9cmFuZDE2KCksIGI9cmFuZDE2KCkpCiAgICAgICAgc2V0UGl4ZWwoeCwgeSwgYykKCmZvciBsaW5lIGluIGV4cG9ydFNjcmVlbigpLnNwbGl0KCdcbicpOgogICAgcHJpbnQobGluZSk=,e}{\color{hyperlinks}{colorful pictures}}.
\item Manipulating the browser's Document Object Model (DOM) to display
  any type of element supported by the browser (text, buttons, menus, etc).
  A basic level of functionality is supported by codeBoot which
  reflects a handful of the JavaScript DOM accessors including
  \texttt{document}, \texttt{querySelector}, \texttt{setAttribute}, and
  \texttt{innerHTML}.
  To allow user input, the DOM elements can have event handling attributes,
  such as \texttt{onclick} and \texttt{onkeypress}, that execute Python
  code, usually a call back to the main program as shown in
  \href{https://codeboot.org/py/?init=A-3jCOlaM3H-SJ3xhYEwsYchQezrRivu4XzrtfqiLF7HHklDtWpvC2kzooifa2mGG88TNjx7xRZd3D_YXj2_PSx9T-hYPWuOj6-Ai_JqS5vhgqZVx3t0r9oLU4wHfnuj,FZXZlbnRzLnB5,d2luID0gZG9jdW1lbnQucXVlcnlTZWxlY3RvcignLmNiLWh0bWwtd2luZG93JykKd2luLmlubmVySFRNTCA9ICc8YnV0dG9uIG9uY2xpY2s9ImNsaWNrKCkiPkFERCAxPC9idXR0b24-JwoKY291bnQgPSAwICAjIGNvdW50IG9mIHRoZSBudW1iZXIgb2YgY2xpY2tzIG9mIHRoZSBidXR0b24KCmRlZiBjbGljaygpOiBnbG9iYWwgY291bnQ7IGNvdW50ICs9IDE7IHByaW50KGNvdW50KQ==,e}{\color{hyperlinks}{Figure~\ref{fig:codeboot-ui}d}}.
\end{itemize}

The ability to access the DOM opens the door to writing standalone web
applications with rich event driven interactions, something we take
advantage of in the last course project.  To develop web applications
that operate outside of the playground area, the codeBoot environment
can be configured into a floating window or be completely hidden to
only show the web application (see
\href{https://codeboot.org/py/?init=K3uQIuX5lhY2MrJsBa7vHglABVUZaWQHDvFP8YuBCvZPekbacxpOB3Y2Bta5xbJSDUzXcEMEePMRhF_WDioKP2QN0oburOVXqQInCURljKnCO9O1BqfigwUGP1umOyu0,hidden:true,floating:true,FdGljLXRhYy10b2UucHk=,IyBGaWNoaWVyOiB0aWMtdGFjLXRvZS5weQoKIyBDZSBwcm9ncmFtbWUgcGVybWV0IMOgIGRldXggcGVyc29ubmVzIGRlIGpvdWVyIGF1IGpldSBkZSB0aWMtdGFjLXRvZS4KIyBMZXMgam91ZXVycywgIngiIGV0ICJvIiwgY2xpcXVlbnQgw6AgbGV1ciB0b3VyIGxhIGNlbGx1bGUgZGUgbGEgZ3JpbGxlIG_DuQojIGlscyBkw6lzaXJlbnQgam91ZXIuICBUcm9pcyBib3V0b25zIHBlcm1ldHRlbnQgZGUgY29tbWVuY2VyIHVuZSBub3V2ZWxsZQojIHBhcnRpZSwgZGUgcmVwcmVuZHJlIHVuIGNvdXAgZXQgZGUgcmVqb3VlciB1biBjb3VwIHJlcHJpcy4KCiMgVmFyaWFibGVzIGdsb2JhbGVzIDoKIwojIHRvdXIgICAgICBlbnRpZXIgaW5kaXF1YW50IMOgIHF1aSBj4oCZZXN0IGxlIHRvdXIsIDEgcG91ciAieCIsIDIgcG91ciAibyIKIyBncmlsbGUgICAgY29udGVudSBkZSBsYSBncmlsbGUgZGUgamV1LCB1biAwIGluZGlxdWUgdW5lIGNlbGx1bGUgdmlkZQojIHJhbmdlZSAgICBjb250ZW51IGRlcyByYW5nZWVzLCB1biBlbnRpZXIgcGFyIHJhbmdlZQojIGNvbG9ubmVzICBjb250ZW51IGRlcyBjb2xvbm5lcywgdW4gZW50aWVyIHBhciBjb2xvbm5lCiMgZGlhZ3MgICAgIGNvbnRlbnUgZGVzIGRpYWdvbmFsZXMKIyBnYWduYW50ICAgdmFsZXVyIGdhZ25hbnRlIHBvdXIgdW5lIHJhbmdlZSwgY29sb25uZSBvdSBkaWFnb25hbGUKIyBvY2N1cGVlcyAgbm9tYnJlIGRlIGNlbGx1bGVzIG9jY3Vww6llcwojIGNvdXBzICAgICB0YWJsZWF1IGRlIHBvc2l0aW9uIGRlcyBjb3VwcyBqb3XDqXMKIyBkaW0gICAgICAgbGFyZ2V1ciBkZSBsYSBncmlsbGUgY2FycsOpZQoKZGltID0gMyAgIyB0aWMtdGFjLXRvZSBzdGFuZGFyZCBzdXIgZ3JpbGxlIDN4MwoKZGVmIHN5bWJvbGVKb3VldXIoam91ZXVyKToKICAgIGlmIGpvdWV1ciA9PSAxOgogICAgICAgIHJldHVybiAnPGltZyBzcmM9Imh0dHBzOi8vY29kZWJvb3Qub3JnL3gucG5nIj4nCiAgICBlbHNlOgogICAgICAgIHJldHVybiAnPGltZyBzcmM9Imh0dHBzOi8vY29kZWJvb3Qub3JnL28ucG5nIj4nCgpkZWYgbm9tSm91ZXVyKGpvdWV1cik6CiAgICByZXR1cm4gJ3gnIGlmIGpvdWV1ciA9PSAxIGVsc2UgJ28nCgpkZWYgYXV0cmVKb3VldXIoam91ZXVyKTogICMgMSAtPiAyLCAyIC0-IDEKICAgIHJldHVybiAzLWpvdWV1cgoKZGVmIGVsZW1lbnQoaWQpOgogICAgcmV0dXJuIGRvY3VtZW50LnF1ZXJ5U2VsZWN0b3IoJyMnICsgaWQpCgpkZWYgY2VsbChpbmRleCk6CiAgICByZXR1cm4gZWxlbWVudCgnY2VsbCcrc3RyKGluZGV4KSkKCmRlZiBhY3RpdmVyKGlkLCBvdWkpOgogICAgaWYgb3VpOgogICAgICAgIGVsZW1lbnQoaWQpLnJlbW92ZUF0dHJpYnV0ZSgnZGlzYWJsZWQnKQogICAgZWxzZToKICAgICAgICBlbGVtZW50KGlkKS5zZXRBdHRyaWJ1dGUoJ2Rpc2FibGVkJywgJ3RydWUnKQoKZGVmIGNsaWNrKGluZGV4KToKICAgIGdsb2JhbCBjb3VwcywgdG91cgogICAgaWYgZ3JpbGxlW2luZGV4XSA9PSAwOgoKICAgICAgICBjb3VwcyA9IGNvdXBzWzpvY2N1cGVlc10KICAgICAgICBjb3Vwcy5hcHBlbmQoaW5kZXgpCgogICAgICAgIGFjdGl2ZXIoJ3VuZG8nLCBUcnVlKQogICAgICAgIGFjdGl2ZXIoJ3JlZG8nLCBGYWxzZSkKCiAgICAgICAgY2VsbChpbmRleCkuaW5uZXJIVE1MID0gc3ltYm9sZUpvdWV1cih0b3VyKQoKICAgICAgICBpZiBham91dGVyR3JpbGxlKGluZGV4LCB0b3VyKToKICAgICAgICAgICAgc2xlZXAoMC4xKSAjIG1ldHRyZSBsZSBkb2N1bWVudCDDoCBqb3VyCiAgICAgICAgICAgIGFsZXJ0KG5vbUpvdWV1cih0b3VyKSArICcgaXMgdGhlIHdpbm5lciEnKQogICAgICAgICAgICBpbml0KCkKICAgICAgICBlbHNlOgogICAgICAgICAgICB0b3VyID0gYXV0cmVKb3VldXIodG91cikKICAgICAgICAgICAgaWYgb2NjdXBlZXMgPT0gZGltKmRpbToKICAgICAgICAgICAgICAgIHNsZWVwKDAuMSkgIyBtZXR0cmUgbGUgZG9jdW1lbnQgw6Agam91cgogICAgICAgICAgICAgICAgYWxlcnQoJ3RoZSBnYW1lIGVuZHMgaW4gYSB0aWUhJykKICAgICAgICAgICAgICAgIGluaXQoKQoKZGVmIHVuZG8oKToKICAgIGdsb2JhbCB0b3VyCiAgICBpZiBvY2N1cGVlcyA-IDA6CiAgICAgICAgY2VsbChjb3Vwc1tvY2N1cGVlcy0xXSkuaW5uZXJIVE1MID0gJycKICAgICAgICByZXRpcmVyR3JpbGxlKGNvdXBzW29jY3VwZWVzLTFdKQogICAgICAgIHRvdXIgPSBhdXRyZUpvdWV1cih0b3VyKQogICAgICAgIGlmIG9jY3VwZWVzID09IDA6CiAgICAgICAgICAgIGFjdGl2ZXIoJ3VuZG8nLCBGYWxzZSkKICAgICAgICBhY3RpdmVyKCdyZWRvJywgVHJ1ZSkKCmRlZiByZWRvKCk6CiAgICBnbG9iYWwgdG91cgogICAgaWYgb2NjdXBlZXMgPCBsZW4oY291cHMpOgogICAgICAgIGNlbGwoY291cHNbb2NjdXBlZXNdKS5pbm5lckhUTUwgPSBzeW1ib2xlSm91ZXVyKHRvdXIpCiAgICAgICAgYWpvdXRlckdyaWxsZShjb3Vwc1tvY2N1cGVlc10sIHRvdXIpCiAgICAgICAgdG91ciA9IGF1dHJlSm91ZXVyKHRvdXIpCiAgICAgICAgaWYgb2NjdXBlZXMgPT0gbGVuKGNvdXBzKToKICAgICAgICAgICAgYWN0aXZlcigncmVkbycsIEZhbHNlKQogICAgICAgIGFjdGl2ZXIoJ3VuZG8nLCBUcnVlKQoKZGVmIGFqb3V0ZXJHcmlsbGUocG9zLCBqb3VldXIpOgogICAgZ2xvYmFsIG9jY3VwZWVzCiAgICB4ID0gcG9zICUgZGltCiAgICB5ID0gcG9zIC8vIGRpbQogICAgZyA9IGdhZ25hbnQgKiBqb3VldXIKICAgIG1hc2tYID0gam91ZXVyIDw8ICgyKngpCiAgICBub3V2Q29sID0gY29sb25uZXNbeF0gfCAoam91ZXVyIDw8ICgyKnkpKQogICAgaWYgbm91dkNvbCA9PSBnOiByZXR1cm4gVHJ1ZQogICAgbm91dlJhbmcgPSByYW5nZWVzW3ldIHwgbWFza1gKICAgIGlmIG5vdXZSYW5nID09IGc6IHJldHVybiBUcnVlCiAgICBub3V2RGlhZzAgPSBkaWFnc1swXSB8IChtYXNrWCBpZiB4PT15IGVsc2UgMCkKICAgIGlmIG5vdXZEaWFnMCA9PSBnOiByZXR1cm4gVHJ1ZQogICAgbm91dkRpYWcxID0gZGlhZ3NbMV0gfCAobWFza1ggaWYgeD09ZGltLTEteSBlbHNlIDApCiAgICBpZiBub3V2RGlhZzEgPT0gZzogcmV0dXJuIFRydWUKICAgIGdyaWxsZVtwb3NdID0gam91ZXVyCiAgICBjb2xvbm5lc1t4XSA9IG5vdXZDb2wKICAgIHJhbmdlZXNbeV0gPSBub3V2UmFuZwogICAgZGlhZ3NbMF0gPSBub3V2RGlhZzAKICAgIGRpYWdzWzFdID0gbm91dkRpYWcxCiAgICBvY2N1cGVlcyArPSAxCiAgICByZXR1cm4gRmFsc2UKCmRlZiByZXRpcmVyR3JpbGxlKHBvcyk6CiAgICBnbG9iYWwgb2NjdXBlZXMKICAgIHggPSBwb3MgJSBkaW0KICAgIHkgPSBwb3MgLy8gZGltCiAgICBtYXNrWCA9IH4oMyA8PCAoMip4KSkKICAgIGdyaWxsZVtwb3NdID0gMAogICAgY29sb25uZXNbeF0gJj0gfigzIDw8ICgyKnkpKQogICAgcmFuZ2Vlc1t5XSAmPSBtYXNrWAogICAgZGlhZ3NbMF0gJj0gKG1hc2tYIGlmIHg9PXkgZWxzZSB-MCkKICAgIGRpYWdzWzFdICY9IChtYXNrWCBpZiB4PT1kaW0tMS15IGVsc2UgfjApCiAgICBvY2N1cGVlcyAtPSAxCgpkZWYgaW5pdEhUTUwoKToKICAgIG1haW4gPSBkb2N1bWVudC5xdWVyeVNlbGVjdG9yKCcjbWFpbicpCiAgICBtYWluLmlubmVySFRNTCA9ICgnJycKICAgICAgPGRpdiBzdHlsZT0iZmxvYXQ6IGxlZnQiPgogICAgICAgIDxicj4KICAgICAgICA8ZGl2PgogICAgICAgICAgPGJ1dHRvbiBvbmNsaWNrPSJpbml0KCkiPk5ldyBnYW1lPC9idXR0b24-CiAgICAgICAgICA8YnV0dG9uIGlkPSJ1bmRvIiBvbmNsaWNrPSJ1bmRvKCkiPlVuZG88L2J1dHRvbj4KICAgICAgICAgIDxidXR0b24gaWQ9InJlZG8iIG9uY2xpY2s9InJlZG8oKSI-UmVkbzwvYnV0dG9uPgogICAgICAgIDwvZGl2PgogICAgICAgIDxicj4KICAgICAgICA8dGFibGU-JycnICsKICAgICAgICAnJy5qb2luKG1hcChsYW1iZGEgeToKICAgICAgICAgICAgICAgICAgICAgICc8dHI-JyArCiAgICAgICAgICAgICAgICAgICAgICAnJy5qb2luKG1hcChsYW1iZGEgeDoKICAgICAgICAgICAgICAgICAgICAgICAgICAgICAgICAgICAgJzx0ZCBpZD0iY2VsbCcgKyBzdHIoeSpkaW0reCkgKwogICAgICAgICAgICAgICAgICAgICAgICAgICAgICAgICAgICAnIiBvbmNsaWNrPSJjbGljaygnICsgc3RyKHkqZGltK3gpICsKICAgICAgICAgICAgICAgICAgICAgICAgICAgICAgICAgICAgJykiPjwvdGQ-JywKICAgICAgICAgICAgICAgICAgICAgICAgICAgICAgICAgIHJhbmdlKGRpbSkpKSArCiAgICAgICAgICAgICAgICAgICAgICAnPC90cj4nLAogICAgICAgICAgICAgICAgICAgIHJhbmdlKGRpbSkpKSArICcnJwogICAgICAgIDwvdGFibGU-CiAgICAgIDwvZGl2PgogICAgJycnKQoKZGVmIGluaXQoKToKICAgIGdsb2JhbCB0b3VyLCBncmlsbGUsIHJhbmdlZXMsIGNvbG9ubmVzLCBkaWFncywgZ2FnbmFudCwgb2NjdXBlZXMsIGNvdXBzCiAgICB0b3VyID0gMQogICAgZ3JpbGxlID0gWzBdICogKGRpbSpkaW0pCiAgICByYW5nZWVzID0gWzBdICogZGltCiAgICBjb2xvbm5lcyA9IFswXSAqIGRpbQogICAgZGlhZ3MgPSBbMF0gKiAyCiAgICBnYWduYW50ID0gKCgxPDwoZGltKjIpKS0xKSAvLyAzCiAgICBvY2N1cGVlcyA9IDAKICAgIGNvdXBzID0gW10KICAgIGluaXRIVE1MKCkKICAgIGZvciBpIGluIHJhbmdlKDkpOgogICAgICAgIGNlbGwoaSkuaW5uZXJIVE1MID0gJycKICAgIGFjdGl2ZXIoJ3JlZG8nLCBGYWxzZSkKICAgIGFjdGl2ZXIoJ3VuZG8nLCBGYWxzZSkKCmluaXQoKQo=,e}{\color{hyperlinks}{Figure~\ref{fig:webapp}}}).
The configuration can be changed using the contextual menu, so the
codeBoot environment can be brought back into view to do more
debugging if needed.

\subsection{Hyperlinks to Execution Snapshots}

Another important pedagogical feature is the creation of hyperlinks
that open codeBoot in the same state (code files, REPL input and point
of execution) as when the program was in originally.  This operation
is available through the contextual menu.  The list of the commands
entered at the REPL, the currently opened files and, if codeBoot is
currently executing a program, the number of steps are all bundled together in a
URL pointing to the codeBoot web site.  Various parts are base64
encoded to satisfy the constraints of URL syntax.  Following the
hyperlink will recreate those REPL interactions and local files, and
if appropriate the execution will be replayed to the same number of
steps.

For security reasons a digital signature of the data is embedded in
the URL and the hyperlink creation feature is only available to
teachers.  This is to prevent students stealing local files from
unsuspecting students by sending them innocuous looking hyperlinks
that execute in their browser to access their files and send them to a
remote location.  Note also that web servers typically have a
(configurable) limit on the length of acceptable URLs.  We have not
encountered this issue with the code examples we use that contain up
to 200 lines of code.

As a demonstration of this feature, hyperlinks were created and
embedded in the current PDF document for various examples including
the previous sections and Figure~\ref{fig:codeboot-ui}.  A simple
click on a screenshot will recreate the example in the browser.
Similar hyperlinks are embedded in the course material (slides, notes,
homework descriptions, web portal, emails, etc).  They greatly improve
the workflow to explain code examples, both for the students, to try
them out and modify them, and the teachers, to switch from the slides
(or other document) to the programming environment quickly.

\subsection{Embedding codeBoot}

For a completely seamless integration with course material, the
codeBoot implementation, a pair of ``\texttt{.js}'' and
``\texttt{.css}'' files, can be included in HTML slides (or other HTML
documents) in such a way as to allow execution and single stepping of
code examples directly from the slides in the web browser.  It is as simple
as wrapping each code example in a \texttt{<pre class="cb-vm">} HTML tag. An HTML page can contain multiple instances of codeBoot which operate independently, making it suitable for documents with sets of examples.

\section{Implementation}
\label{sec:implementation}

The JavaScript and Python interpreters used by codeBoot are not
complete implementations of those languages.  This is acceptable
because a programming course, especially for novices, does not need to
use all the constructs of the language.  As we explain in a later section
the supported constructs go beyond what is absolutely required (for
example Python class definitions, magic methods and exception handling
are all supported even though we don't use them in our course).

The most challenging feature to implement is the single stepping within
the browser environment.  The browser has an execution model
which requires JavaScript code to execute until completion before the
browser handles any events, such as mouse clicks and keypresses, and
refreshes the window, to show any changes to the DOM.  The approach we
used consists of an interpreter in Continuation Passing Style (CPS).

We will focus on the design of the Python interpreter, but the JavaScript
interpreter follows a similar design.

\subsection{Python Interpreter}

The interpreter is based on the \emph{fast interpretation} technique
that transforms the program's Abstract Syntax Tree (AST) into a
function that encapsulates the meaning of that AST.  In a sense this
is a compilation from AST to code represented as a function closure.

\begin{figure*}[th]
  \centering
    \includegraphics[width=6.2in]{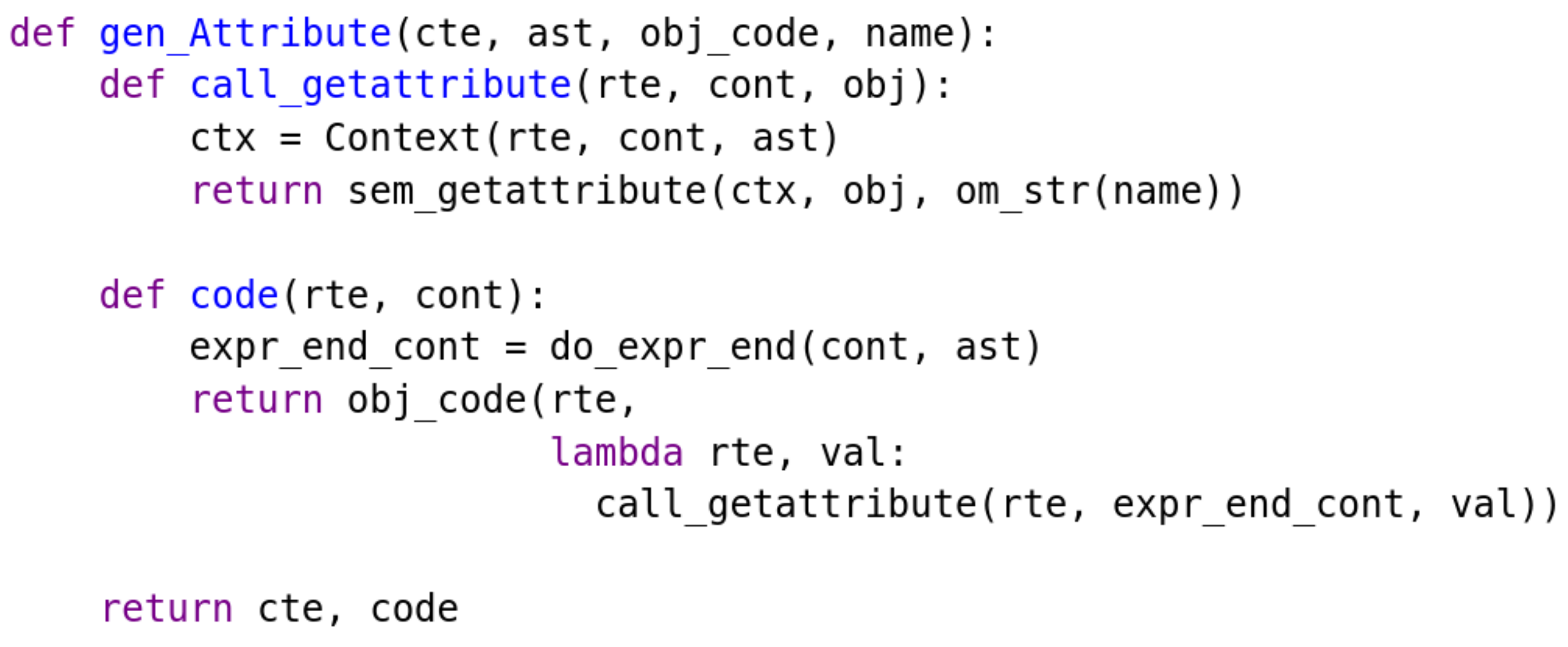}
  \caption{Implementation of the \texttt{\textit{obj}.\textit{attr}} construct in the Python interpreter}
  \label{code:gengetattribute}
\end{figure*}

For another project we had implemented a Python tokenizer and parser
in Python and we decided to reuse it for codeBoot.  This parser
creates ASTs that are compatible with the standard Python \texttt{ast}
module.  Consequently it was natural to prototype our interpreter in
Python and use the standard AST traversal methods.  Once a fairly
complete interpreter was working we wrote a compiler from Python to
JavaScript, \texttt{p2j}, to create a JavaScript version of the
interpreter.  The \texttt{p2j} compiler was relatively
straightforward to write because during development of the interpreter
we had avoided using the more advanced features of Python and the
types and constructs used have a fairly direct mapping to JavaScript.
The features supported by \texttt{p2j} include functions and
closures and the basic types (integers, floating point numbers, and
booleans).  The \texttt{list} and \texttt{str} types are supported but
with few of their standard builtin methods, leaving only the append
and slicing operations. Missing features include most builtin
functions, exception handling and the rest of the Python object model. The
semantics of operators such as \texttt{+} reflects the semantics of
JavaScript.  In comparison our Python interpreter built with
\texttt{p2j} implements a semantics much closer to standard Python than
\texttt{p2j}.

\subsection{CPS and Trampoline}

To implement single-stepping, the interpreter must possess the ability to pause the execution of the code and store its state for later execution. Because the browser's execution model runs JavaScript code until completion, the control flow must be interrupted to let the browser update the window to show the bubble. To allow for pausing, we wrote the interpreter in Continuation Passing Style (CPS). After the execution of any basic operation within an expression (a \textit{step}), the compiled code returns a continuation which takes the form of a JavaScript function. This function is returned to a trampoline, that oversees the chaining of the continuations, along with a special flag which causes the execution to stop. Here, the trampoline has a dual purpose: first, it prevents a stack overflow, since not all browser implement tail-call optimization. Secondly, it stores the continuation and stops the execution when required by the interpreter. When execution must resume, the trampoline is called after restoring the saved execution state.  This will continue execution up to the end of the next step.

In \hyperref[code:gengetattribute]{Figure~\ref{code:gengetattribute}}, we provide the interpreter source code implementing the \texttt{\textit{obj}.\textit{attr}}
construct, which corresponds to an AST of type \texttt{Attribute}. The \texttt{gen\_Attribute} function receives \texttt{cte}, the compile time environment,
\texttt{ast}, the \texttt{Attribute} node (that also contains source code
location information), \texttt{obj\_code}, the code for evaluating
\texttt{\textit{obj}}, and \texttt{name}, the name of
\texttt{\textit{attr}} as a string.  The function returns two values:
the compile time environment (which is unchanged because this node is
not a binding construct), and \texttt{code}, a function that
encapsulates the meaning of the \texttt{\textit{obj}.\textit{attr}}
operation.

The \texttt{code} function takes two parameters.  The first,
\texttt{rte}, is the run time environment which contains the variable
bindings, the current exception handler, the current \texttt{break},
\texttt{continue}, and \texttt{return} destinations, etc.  The second
parameter, \texttt{cont}, is the continuation function indicating
where execution must continue after the code function is done.  The
\texttt{code} function will first evaluate \texttt{\textit{obj}} by
calling \texttt{obj\_code} with the run time environment and the
continuation \texttt{lambda rte, val:...} .  This continuation will
receive the new run time environment and the result of the evaluation.
Finally the Python \texttt{getattribute} operation is executed.  This
operation receives a continuation created by \texttt{do\_expr\_end}
that will (possibly) interact with the UI to show a bubble containing
the information in the run time environment and pointing to the
location indicated by \texttt{ast}. This is done by returning a special
flag to the trampoline to cause it to exit.

\subsection{Python Features Supported by the Interpreter}

Currently the Python interpreter supports the builtin types
\texttt{bool}, \texttt{int}, \texttt{float}, \texttt{str},
\texttt{list}, \texttt{tuple} and \texttt{range} with a few missing
advanced methods.  Other features were not implemented because they
were not necessary for the course, including the types
\texttt{complex}, \texttt{dict} and \texttt{set}, and the constructs
\texttt{del}, \texttt{with}, \texttt{yield}, \texttt{async},
\texttt{await}, list-comprehensions, method decorators, and
type-annotations.

Only a few of the modules of the Python standard library are available
and only a subset of their functions are defined: \texttt{math},
\texttt{random}, \texttt{time}, \texttt{turtle}, and \texttt{functools}.

The set of supported features is sufficient for the development of web
applications of a scale typical of first programming course final
projects.  Large programs can be split into modules and imported
from the browser local file system with the \texttt{import} statement.

\section{Related Work}

There are many online services allowing to create and run code (the list is so long that there is no point listing all systems here). A fair share relies on a remote server for code compilation and execution. This is the case for tryhaskell \cite{doneTryHaskell} (Haskell), Scastie \cite{aleshkaScastie} and ScalaFiddle \cite{scalafiddleScalaFiddle} (Scala). Some also support multiple languages such as Repl.it \cite{repl.itReplIt}, OnlineGDB \cite{geetanjaliOnlineGDB} and Tio \cite{tioTryItOnline}. Remote execution is not desirable as it prevents executing programs which require interaction with the browser, for example when teaching event-driven execution.

Some systems only execute the code within the browser environment: try.scheme.org \cite{schemers.orgTryScheme} and BiwaScheme \cite{haraBiwaScheme} (Scheme), MoonShine \cite{gamesyslimitedMoonShine} (Lua), Try Haxe! \cite{haxecommunityTryHaxe} (Haxe) and CodePen \cite{codepenCodePen} (HTML, CSS and JavaScript). None of the aforementioned implementations support fine-grained single-stepping, with the exception of try.scheme.org, or hyperlink creation. CodePen offers interesting features, but it is specific to JavaScript and does not support single-stepping and hyperlink creation.

We found three mature in-browser Python interpreters: Brython \cite{quentelBrython2012}, Pyodide \cite{iodidePyodide2018} and Skulpt \cite{grahamSkulpt2013}. Despite extensive features, none of them implements fine-grained single-stepping and hyperlink creation.

Environments such as Online Python Tutor \cite{guoOnlinePythonTutor2013} and Pythy \cite{edwardsPythyImprovingIntroductory2014} are specifically aimed at teaching novices. These projects confirmed the benefit of providing an online programming environment to eliminate barriers such as installing the language implementation or a code editor.

Online Python Tutor can create hyperlinks to an exact execution point. It does so by executing the source code on a remote server with the Python debugger module and returning a trace of execution points. This performs well for small programs which limit I/O to the console, but is insufficient for the more complex user interactions needed to teach web programming and event driven execution. Online Python Tutor also limits the execution to 300 steps to guard against excessive long traces, which does not suit the larger scale projects required for our course. Online Python Tutor focuses on visualisation of heap objects contents and pointers. Such a feature is missing from codeBoot and would make a fine addition.

Pythy executes code directly in the browser using a modified version of Skulpt. While it supports line-by-line execution, it does not allow for fine-grained single-stepping nor hyperlink creation.

An alternative to CPS is implemented by Stopify \cite{baxterPuttingAllStops2018}, a JavaScript to JavaScript compiler that performs a transformation to allow pausing or interrupting JavaScript code execution before completion. Stopify aims at making JavaScript a better target for high-level languages. Instead of CPS, the compiler instruments functions such that they have the ability to suspend and resume their own execution. It would be interesting to compare the performance offered by each approach when executing Python in the browser.

\section{Conclusion}

The codeBoot online programming environment was designed to teach
programming to novices.  Students only need a web browser and all
execution is done locally in the browser, allowing students to
continue working while offline and avoiding any special setting up
(such as registering an account or installing softwares).  The fine grained single stepping
feature provided by codeBoot helps students understand the semantics
of the language (priority of operators, flow of control, etc) and the
performance of their code.  The hyperlink creation feature gives
teachers a convenient way to allow students to execute with a single
click the code examples from the course material.  By making key DOM
operations accessible to the program being executed, codeBoot allows
standalone web applications to be written in Python.  To our knowledge
no other online programming environment offers this combination of
features that are valuable for teaching novices.

The codeBoot environment currently implements JavaScript and a subset of the Python
language that is adequate for teaching novices.  We plan to continue
its development to make it even more compatible with the language standards
and more interesting to use for more advanced programming courses. Adding support for other languages is also planned.

The codeBoot source code is available at \mbox{\url{https://github.com/udem-dlteam/codeboot}}.

\vspace*{0.3in}

\noindent\textbf{\large Acknowledgements}

\vspace*{0.15in}

This work was supported by the Natural Sciences and Engineering
Research Council of Canada.  We want to thank the following people who
have helped with the development of codeBoot: Marc-André Bélanger,
Antoine Doucet, Bruno Dufour, Frédéric Hamel, Nicolas Hurtubise,
Léonard Oest O'Leary, and Roselyne Painchaud.

\nocite{*}
\bibliographystyle{eptcs}
\bibliography{tfpie2021}

\appendix

\section*{Appendices}

\begin{figure*}[!hb]
  \centering
  \href{https://codeboot.org/py/?init=K3uQIuX5lhY2MrJsBa7vHglABVUZaWQHDvFP8YuBCvZPekbacxpOB3Y2Bta5xbJSDUzXcEMEePMRhF_WDioKP2QN0oburOVXqQInCURljKnCO9O1BqfigwUGP1umOyu0,hidden:true,floating:true,FdGljLXRhYy10b2UucHk=,IyBGaWNoaWVyOiB0aWMtdGFjLXRvZS5weQoKIyBDZSBwcm9ncmFtbWUgcGVybWV0IMOgIGRldXggcGVyc29ubmVzIGRlIGpvdWVyIGF1IGpldSBkZSB0aWMtdGFjLXRvZS4KIyBMZXMgam91ZXVycywgIngiIGV0ICJvIiwgY2xpcXVlbnQgw6AgbGV1ciB0b3VyIGxhIGNlbGx1bGUgZGUgbGEgZ3JpbGxlIG_DuQojIGlscyBkw6lzaXJlbnQgam91ZXIuICBUcm9pcyBib3V0b25zIHBlcm1ldHRlbnQgZGUgY29tbWVuY2VyIHVuZSBub3V2ZWxsZQojIHBhcnRpZSwgZGUgcmVwcmVuZHJlIHVuIGNvdXAgZXQgZGUgcmVqb3VlciB1biBjb3VwIHJlcHJpcy4KCiMgVmFyaWFibGVzIGdsb2JhbGVzIDoKIwojIHRvdXIgICAgICBlbnRpZXIgaW5kaXF1YW50IMOgIHF1aSBj4oCZZXN0IGxlIHRvdXIsIDEgcG91ciAieCIsIDIgcG91ciAibyIKIyBncmlsbGUgICAgY29udGVudSBkZSBsYSBncmlsbGUgZGUgamV1LCB1biAwIGluZGlxdWUgdW5lIGNlbGx1bGUgdmlkZQojIHJhbmdlZSAgICBjb250ZW51IGRlcyByYW5nZWVzLCB1biBlbnRpZXIgcGFyIHJhbmdlZQojIGNvbG9ubmVzICBjb250ZW51IGRlcyBjb2xvbm5lcywgdW4gZW50aWVyIHBhciBjb2xvbm5lCiMgZGlhZ3MgICAgIGNvbnRlbnUgZGVzIGRpYWdvbmFsZXMKIyBnYWduYW50ICAgdmFsZXVyIGdhZ25hbnRlIHBvdXIgdW5lIHJhbmdlZSwgY29sb25uZSBvdSBkaWFnb25hbGUKIyBvY2N1cGVlcyAgbm9tYnJlIGRlIGNlbGx1bGVzIG9jY3Vww6llcwojIGNvdXBzICAgICB0YWJsZWF1IGRlIHBvc2l0aW9uIGRlcyBjb3VwcyBqb3XDqXMKIyBkaW0gICAgICAgbGFyZ2V1ciBkZSBsYSBncmlsbGUgY2FycsOpZQoKZGltID0gMyAgIyB0aWMtdGFjLXRvZSBzdGFuZGFyZCBzdXIgZ3JpbGxlIDN4MwoKZGVmIHN5bWJvbGVKb3VldXIoam91ZXVyKToKICAgIGlmIGpvdWV1ciA9PSAxOgogICAgICAgIHJldHVybiAnPGltZyBzcmM9Imh0dHBzOi8vY29kZWJvb3Qub3JnL3gucG5nIj4nCiAgICBlbHNlOgogICAgICAgIHJldHVybiAnPGltZyBzcmM9Imh0dHBzOi8vY29kZWJvb3Qub3JnL28ucG5nIj4nCgpkZWYgbm9tSm91ZXVyKGpvdWV1cik6CiAgICByZXR1cm4gJ3gnIGlmIGpvdWV1ciA9PSAxIGVsc2UgJ28nCgpkZWYgYXV0cmVKb3VldXIoam91ZXVyKTogICMgMSAtPiAyLCAyIC0-IDEKICAgIHJldHVybiAzLWpvdWV1cgoKZGVmIGVsZW1lbnQoaWQpOgogICAgcmV0dXJuIGRvY3VtZW50LnF1ZXJ5U2VsZWN0b3IoJyMnICsgaWQpCgpkZWYgY2VsbChpbmRleCk6CiAgICByZXR1cm4gZWxlbWVudCgnY2VsbCcrc3RyKGluZGV4KSkKCmRlZiBhY3RpdmVyKGlkLCBvdWkpOgogICAgaWYgb3VpOgogICAgICAgIGVsZW1lbnQoaWQpLnJlbW92ZUF0dHJpYnV0ZSgnZGlzYWJsZWQnKQogICAgZWxzZToKICAgICAgICBlbGVtZW50KGlkKS5zZXRBdHRyaWJ1dGUoJ2Rpc2FibGVkJywgJ3RydWUnKQoKZGVmIGNsaWNrKGluZGV4KToKICAgIGdsb2JhbCBjb3VwcywgdG91cgogICAgaWYgZ3JpbGxlW2luZGV4XSA9PSAwOgoKICAgICAgICBjb3VwcyA9IGNvdXBzWzpvY2N1cGVlc10KICAgICAgICBjb3Vwcy5hcHBlbmQoaW5kZXgpCgogICAgICAgIGFjdGl2ZXIoJ3VuZG8nLCBUcnVlKQogICAgICAgIGFjdGl2ZXIoJ3JlZG8nLCBGYWxzZSkKCiAgICAgICAgY2VsbChpbmRleCkuaW5uZXJIVE1MID0gc3ltYm9sZUpvdWV1cih0b3VyKQoKICAgICAgICBpZiBham91dGVyR3JpbGxlKGluZGV4LCB0b3VyKToKICAgICAgICAgICAgc2xlZXAoMC4xKSAjIG1ldHRyZSBsZSBkb2N1bWVudCDDoCBqb3VyCiAgICAgICAgICAgIGFsZXJ0KG5vbUpvdWV1cih0b3VyKSArICcgaXMgdGhlIHdpbm5lciEnKQogICAgICAgICAgICBpbml0KCkKICAgICAgICBlbHNlOgogICAgICAgICAgICB0b3VyID0gYXV0cmVKb3VldXIodG91cikKICAgICAgICAgICAgaWYgb2NjdXBlZXMgPT0gZGltKmRpbToKICAgICAgICAgICAgICAgIHNsZWVwKDAuMSkgIyBtZXR0cmUgbGUgZG9jdW1lbnQgw6Agam91cgogICAgICAgICAgICAgICAgYWxlcnQoJ3RoZSBnYW1lIGVuZHMgaW4gYSB0aWUhJykKICAgICAgICAgICAgICAgIGluaXQoKQoKZGVmIHVuZG8oKToKICAgIGdsb2JhbCB0b3VyCiAgICBpZiBvY2N1cGVlcyA-IDA6CiAgICAgICAgY2VsbChjb3Vwc1tvY2N1cGVlcy0xXSkuaW5uZXJIVE1MID0gJycKICAgICAgICByZXRpcmVyR3JpbGxlKGNvdXBzW29jY3VwZWVzLTFdKQogICAgICAgIHRvdXIgPSBhdXRyZUpvdWV1cih0b3VyKQogICAgICAgIGlmIG9jY3VwZWVzID09IDA6CiAgICAgICAgICAgIGFjdGl2ZXIoJ3VuZG8nLCBGYWxzZSkKICAgICAgICBhY3RpdmVyKCdyZWRvJywgVHJ1ZSkKCmRlZiByZWRvKCk6CiAgICBnbG9iYWwgdG91cgogICAgaWYgb2NjdXBlZXMgPCBsZW4oY291cHMpOgogICAgICAgIGNlbGwoY291cHNbb2NjdXBlZXNdKS5pbm5lckhUTUwgPSBzeW1ib2xlSm91ZXVyKHRvdXIpCiAgICAgICAgYWpvdXRlckdyaWxsZShjb3Vwc1tvY2N1cGVlc10sIHRvdXIpCiAgICAgICAgdG91ciA9IGF1dHJlSm91ZXVyKHRvdXIpCiAgICAgICAgaWYgb2NjdXBlZXMgPT0gbGVuKGNvdXBzKToKICAgICAgICAgICAgYWN0aXZlcigncmVkbycsIEZhbHNlKQogICAgICAgIGFjdGl2ZXIoJ3VuZG8nLCBUcnVlKQoKZGVmIGFqb3V0ZXJHcmlsbGUocG9zLCBqb3VldXIpOgogICAgZ2xvYmFsIG9jY3VwZWVzCiAgICB4ID0gcG9zICUgZGltCiAgICB5ID0gcG9zIC8vIGRpbQogICAgZyA9IGdhZ25hbnQgKiBqb3VldXIKICAgIG1hc2tYID0gam91ZXVyIDw8ICgyKngpCiAgICBub3V2Q29sID0gY29sb25uZXNbeF0gfCAoam91ZXVyIDw8ICgyKnkpKQogICAgaWYgbm91dkNvbCA9PSBnOiByZXR1cm4gVHJ1ZQogICAgbm91dlJhbmcgPSByYW5nZWVzW3ldIHwgbWFza1gKICAgIGlmIG5vdXZSYW5nID09IGc6IHJldHVybiBUcnVlCiAgICBub3V2RGlhZzAgPSBkaWFnc1swXSB8IChtYXNrWCBpZiB4PT15IGVsc2UgMCkKICAgIGlmIG5vdXZEaWFnMCA9PSBnOiByZXR1cm4gVHJ1ZQogICAgbm91dkRpYWcxID0gZGlhZ3NbMV0gfCAobWFza1ggaWYgeD09ZGltLTEteSBlbHNlIDApCiAgICBpZiBub3V2RGlhZzEgPT0gZzogcmV0dXJuIFRydWUKICAgIGdyaWxsZVtwb3NdID0gam91ZXVyCiAgICBjb2xvbm5lc1t4XSA9IG5vdXZDb2wKICAgIHJhbmdlZXNbeV0gPSBub3V2UmFuZwogICAgZGlhZ3NbMF0gPSBub3V2RGlhZzAKICAgIGRpYWdzWzFdID0gbm91dkRpYWcxCiAgICBvY2N1cGVlcyArPSAxCiAgICByZXR1cm4gRmFsc2UKCmRlZiByZXRpcmVyR3JpbGxlKHBvcyk6CiAgICBnbG9iYWwgb2NjdXBlZXMKICAgIHggPSBwb3MgJSBkaW0KICAgIHkgPSBwb3MgLy8gZGltCiAgICBtYXNrWCA9IH4oMyA8PCAoMip4KSkKICAgIGdyaWxsZVtwb3NdID0gMAogICAgY29sb25uZXNbeF0gJj0gfigzIDw8ICgyKnkpKQogICAgcmFuZ2Vlc1t5XSAmPSBtYXNrWAogICAgZGlhZ3NbMF0gJj0gKG1hc2tYIGlmIHg9PXkgZWxzZSB-MCkKICAgIGRpYWdzWzFdICY9IChtYXNrWCBpZiB4PT1kaW0tMS15IGVsc2UgfjApCiAgICBvY2N1cGVlcyAtPSAxCgpkZWYgaW5pdEhUTUwoKToKICAgIG1haW4gPSBkb2N1bWVudC5xdWVyeVNlbGVjdG9yKCcjbWFpbicpCiAgICBtYWluLmlubmVySFRNTCA9ICgnJycKICAgICAgPGRpdiBzdHlsZT0iZmxvYXQ6IGxlZnQiPgogICAgICAgIDxicj4KICAgICAgICA8ZGl2PgogICAgICAgICAgPGJ1dHRvbiBvbmNsaWNrPSJpbml0KCkiPk5ldyBnYW1lPC9idXR0b24-CiAgICAgICAgICA8YnV0dG9uIGlkPSJ1bmRvIiBvbmNsaWNrPSJ1bmRvKCkiPlVuZG88L2J1dHRvbj4KICAgICAgICAgIDxidXR0b24gaWQ9InJlZG8iIG9uY2xpY2s9InJlZG8oKSI-UmVkbzwvYnV0dG9uPgogICAgICAgIDwvZGl2PgogICAgICAgIDxicj4KICAgICAgICA8dGFibGU-JycnICsKICAgICAgICAnJy5qb2luKG1hcChsYW1iZGEgeToKICAgICAgICAgICAgICAgICAgICAgICc8dHI-JyArCiAgICAgICAgICAgICAgICAgICAgICAnJy5qb2luKG1hcChsYW1iZGEgeDoKICAgICAgICAgICAgICAgICAgICAgICAgICAgICAgICAgICAgJzx0ZCBpZD0iY2VsbCcgKyBzdHIoeSpkaW0reCkgKwogICAgICAgICAgICAgICAgICAgICAgICAgICAgICAgICAgICAnIiBvbmNsaWNrPSJjbGljaygnICsgc3RyKHkqZGltK3gpICsKICAgICAgICAgICAgICAgICAgICAgICAgICAgICAgICAgICAgJykiPjwvdGQ-JywKICAgICAgICAgICAgICAgICAgICAgICAgICAgICAgICAgIHJhbmdlKGRpbSkpKSArCiAgICAgICAgICAgICAgICAgICAgICAnPC90cj4nLAogICAgICAgICAgICAgICAgICAgIHJhbmdlKGRpbSkpKSArICcnJwogICAgICAgIDwvdGFibGU-CiAgICAgIDwvZGl2PgogICAgJycnKQoKZGVmIGluaXQoKToKICAgIGdsb2JhbCB0b3VyLCBncmlsbGUsIHJhbmdlZXMsIGNvbG9ubmVzLCBkaWFncywgZ2FnbmFudCwgb2NjdXBlZXMsIGNvdXBzCiAgICB0b3VyID0gMQogICAgZ3JpbGxlID0gWzBdICogKGRpbSpkaW0pCiAgICByYW5nZWVzID0gWzBdICogZGltCiAgICBjb2xvbm5lcyA9IFswXSAqIGRpbQogICAgZGlhZ3MgPSBbMF0gKiAyCiAgICBnYWduYW50ID0gKCgxPDwoZGltKjIpKS0xKSAvLyAzCiAgICBvY2N1cGVlcyA9IDAKICAgIGNvdXBzID0gW10KICAgIGluaXRIVE1MKCkKICAgIGZvciBpIGluIHJhbmdlKDkpOgogICAgICAgIGNlbGwoaSkuaW5uZXJIVE1MID0gJycKICAgIGFjdGl2ZXIoJ3JlZG8nLCBGYWxzZSkKICAgIGFjdGl2ZXIoJ3VuZG8nLCBGYWxzZSkKCmluaXQoKQo=,e}{\includegraphics[width=3.7in]{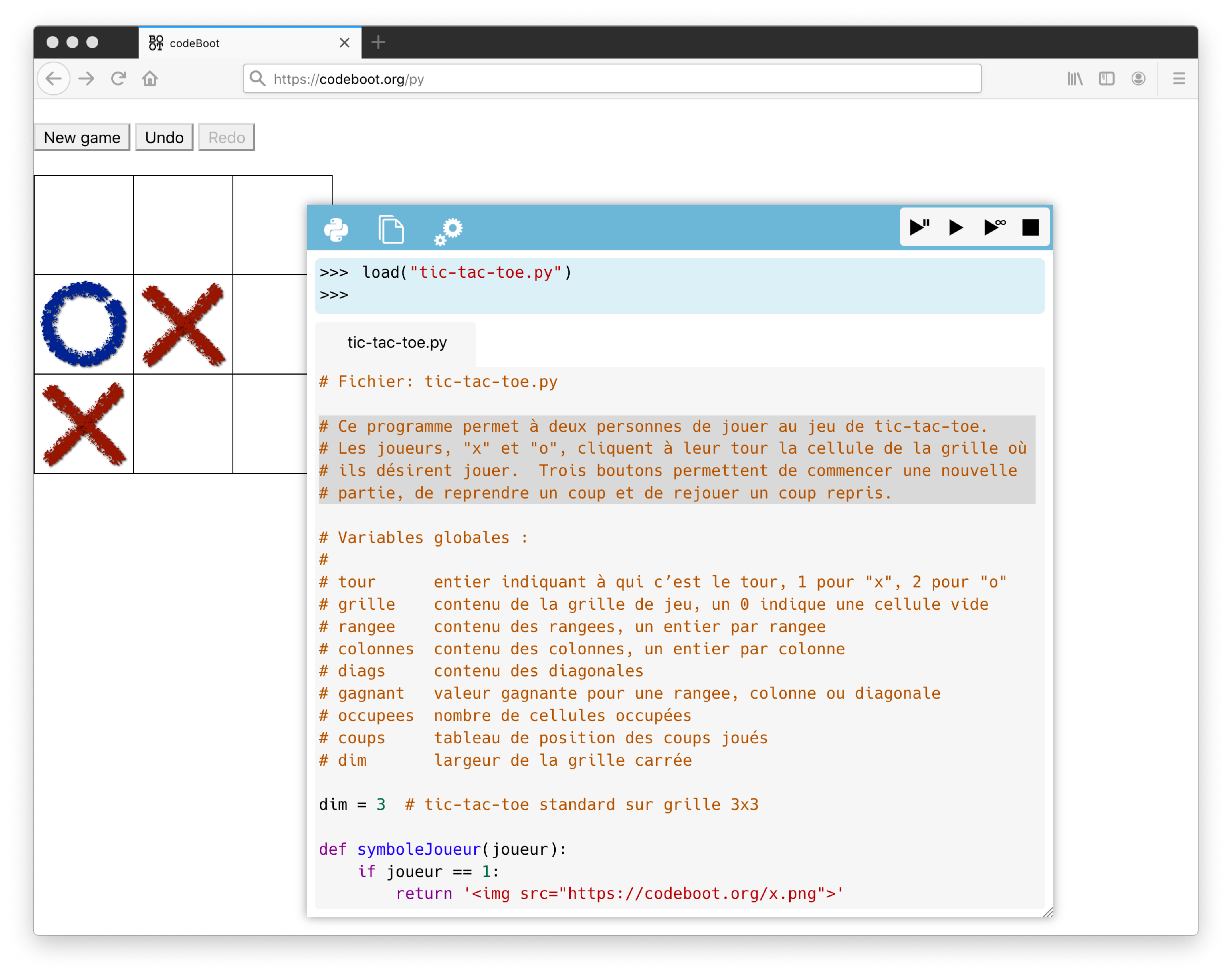}}
  \caption{Using codeBoot, programs written in Python can be bundled as web applications in which the codeBoot environment can be hidden or brought into view for debugging}\label{fig:webapp}
\vspace{.5cm}
\centering
\includegraphics[width=3.5in]{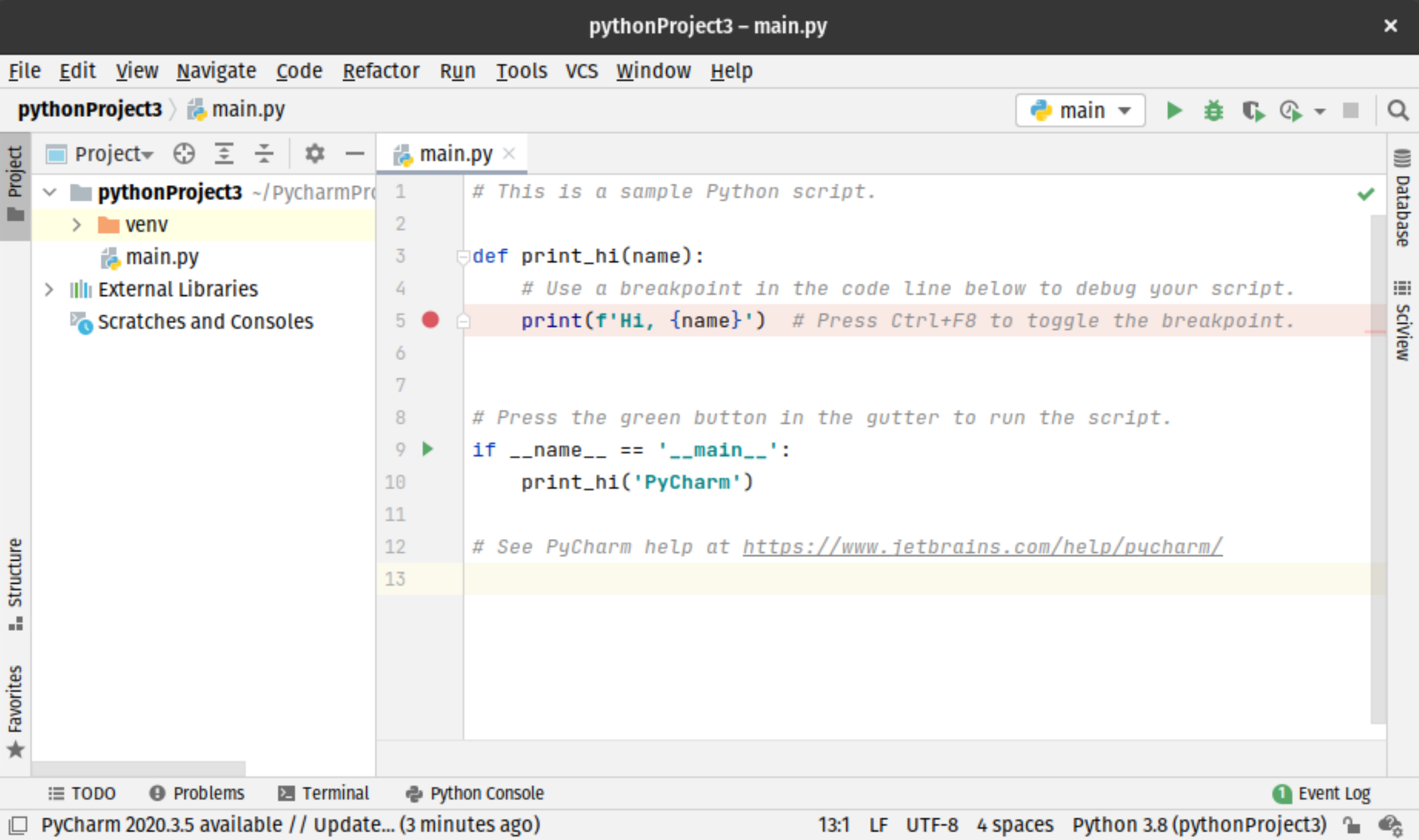}
\caption{User interface of the PyCharm IDE}
\label{design:pycharm}
\vspace*{-2.0in}
\end{figure*}

\end{document}